\documentclass[12pt]{article}
\usepackage{amsfonts}

\begin{document}

\def\ba{\begin{eqnarray}}
\def\ea{\end{eqnarray}}
\def\w{\wedge}

\begin{titlepage}
\title{{\bf  String-Inspired Chern-Simons Modified Gravity In 4-Dimensions}}
\author{ M. Adak\footnote
{madak@pau.edu.tr}\\
 {\small Department of Physics, Pamukkale University}\\{\small 20017 Denizli, Turkey} \\  \\
T. Dereli\footnote{ tdereli@ku.edu.tr} \\
 {\small Department of Physics, Ko\c{c} University}\\
 {\small 34450 Sar{\i}yer-\.{I}stanbul, Turkey  }  \\
} \vskip 1cm
\date{{05 January 2012}, {\it file StringChernSimonsGravity03.tex} }
\maketitle

\begin{abstract}

\noindent Field equations of the Chern-Simons modified gravity in
4-dimensions are obtained by a truncation of the field equations of
the low energy effective string models  with first order corrections
in the string constant included.

\vskip 1cm

\end{abstract}
\end{titlepage}

\section{Introduction}

\noindent A Chern-Simons modified gravity in $D=4$ dimensions
proposed by Jackiw and Pi \cite{jackiw2003,jackiw2004} has
attracted a lot of attention recently \cite{alexander2009}. The
Jackiw-Pi model is derived from an action that consists of the
usual Einstein-Hilbert term plus a topological term with a cosmic
scalar field $\theta$ appearing as a Lagrange multiplier. It was
shown in the linearized approximation that one of two polarization
states of the graviton is suppressed due to this modification. It
was also noticed that the Kerr solution of general relativity is
forbidden in this theory because of the zero-Pontryagin
constraint. This  point was later analyzed in detail by Grumiller
and Yunes \cite{grumiller2008}. In a more recent study, the
effects of a slightly extended model on bodies orbiting the Earth
were discussed by Smith et al \cite{smith2008}. It was suggested
by Satoh, Kanno and Soda \cite{satoh2008} that such effects should
be viable in a string-inspired inflationary cosmology. An
Einstein-Cartan approach leading to an interpretation with
dynamical torsion was also found \cite{alexander2008,
cantcheff2008}. The interest on the 4D Chern-Simons gravity still
continues. For example, as a slowly rotating black hole reducing
to the Kerr solution \cite{kkonno2009} and perturbations of
spherically symmetric black hole on the assumption of vanishing
background scalar field \cite{cmolina2010} were studied, in
Ref\cite{hahmedov2010} the decoupling and reduction properties
were analyzed. In this context it is our wish here to point out
that the Jackiw-Pi model of Chern-Simons modified gravity and its
extensions above can be accommodated within the framework of
effective string field theory in $D=4$ dimensions with first order
corrections in string parameter $\alpha'$ taken into account.
Furthermore, we obtained the pp-wave solution to the effective
string field theory for special choice of constant dilaton and
noticed that this solution is indistinguishable from general
relativity in the Jackiw-Pi model. The notation and further
details may be found in an older paper by one of us
\cite{dereli1987}.

\section{Effective String Theory Field Equations}

\noindent We will start by examining the bosonic field equations
arising from the action $I = \int_M L$ where $M$ is a 4-dimensional
manifold and the 4-form $L$ is given in terms of a dilaton 0-form
$\phi$, and a 3-form field $H$. We take an action
 \ba
    L_0 = e^{-\phi} \left ( R_{ab} \w *e^{ab} -\alpha d\phi \w *d\phi + \beta H \w *H + \lambda
    *1 \right )
 \ea
 to the zeroth order in
$\alpha'$. The metric here has been scaled in accordance with a
heterotic string model. But the conclusions will not be changed in
principle if we adopt a scaling appropriate to different string
models. We employ Lorentzian connection 1-forms $\omega^a{}_b$ in
terms of which $R^a{}_b= d\omega^a{}_b + \omega^a{}_c \w
\omega^c{}_b$. The metric tensor $g$ on $M$ is given in terms of an
orthonormal coframe $\{ e^a \}$ by $g=\eta_{ab} e^a \otimes e^b$
with $\eta_{ab}= \mbox{diag}(-+++)$. The space-time orientation is
fixed by the volume 4-form $*1 = e^0 \w e^1 \w e^2 \w e^3$. The
constants $\alpha , \beta$ are order zero in $\alpha'$ while
$\lambda$ is of order ${\alpha'}^{-1}$. We at first set the
connection to be the unique metric compatible, torsion-free
Levi-Civita connection. We impose this assumption by introducing the
constraint term
 \ba
     L_C = (de^a + \omega^a{}_b \w e^b) \w \lambda_a
 \ea
in the action  where $\lambda_a$ are Lagrange multiplier 2-forms. We
will comment on the possibility of having  dynamical torsion later.
 An essential contribution at first order in $\alpha'$ to $L$ comes
from the Lorentz Chern-Simons form necessary for anomaly cancelation
in string models. Other quadratic contributions of the curvature
tensor such as the Euler-Poincar\'{e} density are also required. We
will repeat our previous analysis in Ref.\cite{dereli1987} in $D=4$
dimensions with
 \ba
    L_1 = \eta R_{ab} \w R_{cd} *e^{abcd} + \mu (dH - \epsilon R_{ab} \w
    R^{ab}) .
 \ea
The constants $\eta , \epsilon$ are  first order in string parameter
$\alpha'$ and $\mu$ is a Lagrange multiplier 0-form that enforces
the Bianchi condition $dH = \epsilon R_{ab} \w R^{ab}$. It is
well-known that $R_{ab} \w R^{ab} = dK $  where the Chern-Simons
3-form $K=\omega_{ab} \w d\omega^{ab} + \frac{2}{3} \omega_{ab} \w
\omega^a{}_c \w \omega^{cb}$. In fact,  Euler-Poincar\'{e} density
$R_{ab} \w R_{cd}
*e^{abcd}$ is also an exact form in $D=4$ dimensions. We keep it in the action but it
doesn't give any contribution to the variational field equations.

\bigskip

\noindent In order to derive the effective string theory field
equations, the action $I=\int_M (L_0 + L_1 + L_C) $ is going to be
varied as a functional of the variables $\{ e^a , \phi , H,
\omega^a{}_b \}$ in a fixed local coordinate chart for $M$, subject
to zero-torsion constraint. The Einstein field equations from the
coframe $e^a$-variations are
 \ba
     e^{-\phi} \left ( G_a + \alpha \tau_a[\phi] -\beta \tau_a[H] +
     \lambda *e_a \right ) + D\lambda_a = 0  \label{eq:Einstein0}
 \ea
where the Einstein 3-forms \ba
 G_a = R^{bc} \wedge *e_{abc} ,
\ea the dilaton stress-energy 3-forms \ba \tau_a[\phi] =  \iota_a
d\phi *d\phi + d\phi \wedge \iota_a *d\phi , \ea and the axion
stress-energy 3-forms \ba \tau_a[H] =  \iota_a H \wedge
*H + H \w \iota_a *H  .
\ea The dilaton field equation from the  $\phi$-variation is
 \ba
   e^{-\phi} (R_{ab} \w *e^{ab} - \alpha d\phi \w *d\phi + \beta H \w *H)
   - 2 \alpha d(e^{-\phi} *d\phi) + \lambda e^{-\phi} *1 =0 .
   \label{eq:dilatoneqn0}
 \ea
$H$-variation gives
 \ba
  2\beta e^{-\phi} *H + d\mu =0
 \ea
while from the $\mu$-variation, the constraint
 \ba
 dH = \epsilon R_{ab} \w R^{ab} \label{eq:constraint}
 \ea
is obtained. Since $d^2\mu =0$, the $H$-field equation may be
replaced by
 \ba
   \beta d(e^{-\phi} *H) = 0 . \label{eq:H-variation}
 \ea
We also  vary the  connection 1-forms $\omega^a{}_b$ as independent
variables and  obtain
 \ba
   - e^{-\phi} d\phi \w *e_{ab} + 4\epsilon \beta e^{-\phi} R_{ab}
   \w *H = \frac{1}{2}(e_a \w \lambda_b - e_b \w \lambda_a).
 \ea
These are algebraic for the multiplier 2-forms and admit the unique
solution
  \ba
    \lambda_a = -2 e^{-\phi} \imath_a *d\phi + 8\epsilon \beta e^{-\phi} \imath^b
    (R_{ba} \w *H) + 2\epsilon \beta e_a \w e^{-\phi} \imath^b
    \imath^c  (R_{bc} \w *H) .
 \ea
Substituting this back into the Einstein field equations
(\ref{eq:Einstein0}) we get
 \ba
  & & e^{-\phi} (G_a + \alpha \tau_a[\phi] -\beta \tau_a[H] +
     \lambda *e_a )  -2 D(e^{-\phi} \imath_a *d\phi) \nonumber \\
   & &  + 8\epsilon \beta D(e^{-\phi} \imath^b
    (R_{ba} \w *H)) - 2\epsilon \beta e_a \w D(e^{-\phi} \imath^b
    \imath^c  (R_{bc} \w *H)) = 0 .\label{eq:Einstein1}
 \ea
Taking the trace of it and comparing with (\ref{eq:dilatoneqn0}), we
see that the dilaton field equation (\ref{eq:dilatoneqn0}) may be
replaced by
 \ba
     (3+2\alpha) d(e^{-\phi} *d\phi) = 2 \beta e^{-\phi} H \wedge
     *H - \lambda e^{-\phi} *1 . \label{eq:dilaton}
 \ea
The coupled system of field equations (\ref{eq:constraint}),
(\ref{eq:H-variation}), (\ref{eq:Einstein1}) and (\ref{eq:dilaton})
describes an axi-dilaton gravity theory that includes Jackiw-Pi
model and its extensions as special cases. In order to prove this
claim,  we  consider a special class of solutions with constant
dilaton: $\phi = \phi_0$.  The reduced field equations become
 \ba
     G_a - \beta \tau_a[H] +
     \lambda *e_a  + 8\epsilon \beta D(\imath^b
    (R_{ba} \w *H)) & & \nonumber \\
     - 2\epsilon \beta e_a \w D( \imath^b
    \imath^c  (R_{bc} \w *H)) &=& 0 , \\
    2 \beta H \w *H - \lambda *1 &=& 0 , \\
  dH - \epsilon R_{ab} \w R^{ab} &=& 0 ,\\
  \beta (d*H ) &=& 0 .
 \ea
It is always possible to write locally in 4-dimensions,
 \ba
     *H = d\theta .
 \ea
In terms of the axion 0-form $\theta$ the above reduced set of field
equations reads
  \ba
     G_a - \beta \tau_a[\theta] + \lambda *e_a  + 8\epsilon \beta D(\imath^b (R_{ba} \w d\theta)) & & \nonumber \\
     - 2\epsilon \beta e_a \w D( \imath^b \imath^c  (R_{bc} \w d\theta)) &=& 0 , \label{eq:modeinseqn}\\
    2 \beta d\theta  \w *d\theta + \lambda *1 &=& 0 , \label{eq:lambda}\\
  \beta d*d\theta - \epsilon \beta R_{ab} \w R^{ab} &=& 0 . \label{eq:R2}
 \ea
These are the field equations of Smith et al \cite{smith2008} with a
judicious choice of coupling constants. We may further truncate the
system by letting $\beta \rightarrow 0$ and $\lambda \rightarrow 0$
while keeping $\epsilon \beta =1$. Then we arrive at
  \ba
     G_a + 8 D(\imath^b (R_{ba} \w d\theta)) - 2 e_a \w D( \imath^b \imath^c  (R_{bc} \w d\theta)) &=& 0 , \label{eq:Jac1}\\
   R_{ab} \w R^{ab} &=& 0 \label{eq:Jac2}
 \ea
that are precisely the field equations of Jackiw and Pi
\cite{jackiw2003}.

\section{Concluding Comments}

\noindent We first find pp-wave solution to the Eqn.s
(\ref{eq:modeinseqn})-(\ref{eq:R2})
 \ba
     g= -c^2 dt^2 + p^2(u) dx^2 + q^2(u) dy^2 + dz^2 \quad , \quad \theta = \theta(u)
 \ea
where $u=z-ct$ is null coordinate. For such metrics, the Pontryagin density 4-form vanishes identically, $R_{ab} \w R^{ab}=0$. Thus CS field $\theta(u)$ satisfies $d*d\theta=0$ in accordance with (\ref{eq:R2}). Also because of the result $d\theta  \w *d\theta =0$, which means that CS velocity field $v_a := \partial_a \theta$ is a null Killing vector $v^av_a=0$, we arrive at $\lambda =0$ via (\ref{eq:lambda}). We note that the last two terms on the left hand side of (\ref{eq:modeinseqn}), the so-called C-tensor in Ref\cite{alexander2009}, are zero. Finally we calculate the Einstein 3-forms and the energy-momentum 3-forms of the CS field as
 \ba
   G_0 = -G_3 = 2\left( \frac{p''}{p} + \frac{q''}{q} \right) e^{12} \w (e^0 - e^3) \; , \quad G_1=G_2=0 \; , \\
   \tau_0 [\theta] = -\tau_3[\theta] = 2 \left( \theta' \right)^2 e^{12} \w (e^0 -e^3) \; , \quad \tau_1 [\theta] = \tau_2[\theta] =0
 \ea
where prime denotes derivative with respect to $u$. Thus we end up the modified Einstein equation
 \ba
    \frac{p''}{p} + \frac{q''}{q} = \beta \left( \theta' \right)^2 \, .\label{eq:modeinseqn2}
 \ea
In much of the work on Chern-Simons gravity, the Cern-Simons term
is coupled to a scalar field (we did the same indirectly), and
this scalar field is assumed to be spatially homogeneous but time
varying, $\theta = \mu t$ where $\mu$ any constant and $t$ time
coordinate, (we did not do the same). We found that in the empty
space-time gravitational waves still have two polarizations and
propagate with speed of light but have explicitly modified
profiles via (\ref{eq:modeinseqn2}). We also want to remark that
Jackiw-Pi model given by (\ref{eq:Jac1}) and (\ref{eq:Jac2}) for
the choice $\theta = \mu t$ as assumed by themselves is
indistinguishable from general relativity in the context of
pp-waves.

 \bigskip

\noindent Secondly we comment on variational field equations with
dynamical torsion. In this case, independent
$\omega^a{}_b$-variations of the action $\int_M (L_0 + L_1)$ yield
the field equations
 \ba
    e^{-\phi} \left (  T^a \w *e_{abc} - \frac{1}{2} d\phi \w e^a \w *e_{abc} + 4 \epsilon \beta
    R_{ab} \w *H \right )  = 0 . \label{torsion}
 \ea
  These are not  algebraic
for the connection owing to the presence of curvature forms. We thus
encounter here propagating torsion. In fact, connection 1-forms with
torsion are uniquely decomposed according to
 \ba
    \omega^a{}_b =   \hat{\omega}^a{}_b + K^a{}_b
 \ea
where $\hat{\omega}^a{}_b$ are the Levi-Civita connection 1-forms
and  $K^a{}_b$ are the contortion 1-forms such that $ K^a{}_b \w e^b
= T^a  $. Then the corresponding curvature 2-forms are similarly
decomposed as
 \ba
    R^a{}_b  = \hat{R}^a{}_b + \hat{D} K^a{}_b +  K^a{}_c \w K^c{}_b
 \ea
so that (\ref{torsion}) becomes a highly non-linear set of first
order differential equations in the components of $T^a$'s. For
further details we refer to \cite{dereli1987}.

\bigskip

\noindent To summarize, the axi-dilaton gravity in $D=4$
dimensions described by the field equations (\ref{eq:constraint}),
(\ref{eq:H-variation}), (\ref{eq:Einstein1}) and
(\ref{eq:dilaton}) includes   the Chern-Simons modified gravity of
Jackiw and Pi and its extensions that appeared in recent
literature as  special cases. Axi-dilaton gravity is a
self-consistent theory of gravity whose solutions besides pp-waves
obtained above both in $D=4$ \cite{dereli1995}  and $D\geq 5$
\cite{dereli2007} dimensions are worthy of study in their own
right from the physical view point.

\end{document}